\documentclass[aps,prd,tighten,showpacs,a4]{revtex4}

\usepackage{graphicx}
\usepackage{bm}
\usepackage{amsmath,amssymb}
\usepackage{latexsym}

\newcommand{\rp}{r_{+}}

\newcommand{\rc}{\tilde{r}_{c}}

\newcommand{\pp}{\phantom}
\newcommand{\tT}{\tilde{T}}

\begin{document}

\title{Semiclassical lukewarm black holes}

\author{Jerzy Matyjasek and Katarzyna Zwierzchowska}
\affiliation{Institute of Physics,
Maria Curie-
Sk\l odowska University\\
pl. Marii Curie-
Sk\l odowskiej 1,
20-031 Lublin, Poland}

\date{\today}

\begin{abstract}
The perturbative solutions to the semiclassical Einstein field equations 
describing spherically-symmetric and static lukewarm black 
hole are constructed. The source term is composed of 
the (classical) stress-energy tensor of the electromagnetic field and 
the renormalized stress-energy tensor of the quantized massive 
scalar field in a large mass limit. We used two different parametrizations. 
In the first parametrization we calculated the zeroth-order solution.
Subsequently, making use of the quantum part of the total stress-energy tensor
constructed in the classical background we calculated the corrections to the
metric potentials and the corrections to the horizons. This procedure 
can be thought of as switching the quantized field on and analyzing its
influence on the classical background via the back-reaction.
In the second parametrization we are looking for
a self-consistent lukewarm solution from the very beginning. 
This requires knowledge of a generic tensor which depend functionally on the metric tensor. 
The transformation formulas relating the line element in both parametrizations
are given. 

\end{abstract}



\maketitle
\section{Introduction}


One of the most characteristic features of the Riessner-Nordstr\"om-
deSitter black holes is the simultaneous occurrence of the three
horizon-like surfaces: the inner horizon, the event horizon and the 
cosmological horizon. This reflects the fact that the equation $g_{tt}(r) =0$ 
has four real roots (not necessarily distinct), three of which are positive
and represent horizons whereas the negative root has no physical interpretation. 
This leads to a number of special cases with the near-horizon geometries
described by the Bertotti-Robinson~\cite{bertotti,robinson}, Nariai~\cite{Nariai1,
Nariai2} or  Plebanski-Hacyan~\cite{Hacyan} line elements.
Among various allowable black hole configurations the class of solutions in which 
the temperature of the event horizon equals the temperature of the cosmological
horizon is special~\cite{Romans,Mellor1,Mellor2}. 
It is because the mean values of the characteristics
of the quantized fields, such as the field fluctuation and 
the renormalized stress-energy tensor are expected to be regular 
in the thermal state at the natural temperature. These expectations
have been confirmed by a direct calculation of the stress-energy tensor 
in the two dimensional setting~\cite{Lizzie1}  and the field fluctuation of the massive
quantized scalar field in the full four dimensional geometry~\cite{Lizzie2}. 
The black holes for which both temperatures are equal are usually referred to 
as the lukewarm black holes. Recent analyses include~\cite{Breen,Kaska3,
Dunajski}. It should be noted that there is no lukewarm configuration 
for the Schwarzschild-de Sitter black hole. 

Since the stress-energy tensor of the quantized fields contribute to the
source term of the semiclassical Einstein field equations, the resulting
geometry changes due to the back reaction process. Consequently, the natural 
question arises: Whether or not it is possible to construct the lukewarm 
black hole in the semiclassical gravity. In Ref~\cite{Kaska3} a related problem
has been considered in the context of the (classical) quadratic gravity.
Specifically, it has been shown, that although the full, detailed answer 
is beyond our capabilities, it is possible to provide an affirmative answer 
to the restricted problem in which the differential equations describing the model
are solved perturbatively. These results can also be viewed as a first step
towards incorporation of the quantum effects into the picture. It is because 
the renormalized stress-energy tensor of the quantized massive field in the large
mass limit  may be  approximated by the object constructed from the curvature 
tensor, its covariant derivatives and contractions. In this approach one ignores
the particle creation which is a nonlocal phenomenon. In spite of this limitation
this framework is still the most general one not restricted to any particular 
type of symmetry.   

Here we generalize the results of Ref.~\cite{Kaska3} in a twofold way: 
First, we employ the renormalized stress-energy tensor of the massive
scalar field with the arbitrary curvature coupling constructed within 
the framework of the Schwinger-DeWitt approximation to construct the 
semiclassical lukewarm black hole. Secondly, we study the relation
between the results expressed in terms of the radii of the cosmological and
event horizons of the classical lukewarm solution 
and the analogous results constructed in $(\rp,\rc)$ parametrization,
where $\rp$ is the exact location of the event horizon of the semiclassical
black hole and $\rc$ describes the cosmological horizon in a zeroth-order 
approximation.     

A few words on the method are in order. First observe that first three
terms of the Schwinger-DeWitt expansion are divergent and can be absorbed into 
the (classical) gravitational action with the cosmological constant and the
quadratic terms~\cite{Birrell}. This renormalizations leaves us with the two additional
parameters, say, $\alpha$ and $\beta$ which should be determined observationally.
Their exact values are presently unknown, it is expected, however, that they 
are small, otherwise they would lead to various observational effects.
To simplify calculations, in what follows, we shall set them to zero.
The second observation is related to the perturbative approach in the  effective 
theories. In fact it may be the only method to deal with them. Indeed, since 
the semiclassical gravity involves six-order derivatives of the metric 
their nonperturbative solutions may appear to be spurious and one has 
to invent a method for systematic selecting physical ones. It seems
that the acceptable solutions, when expanded in powers of the small 
parameter, should reduce to those obtained within the framework of 
perturbative approach. 
Finally observe that there are good reasons to believe that (in most cases)
the black hole exists
as perturbative solution of the higher order equations provided it 
exists classically~\cite{RCM1}.

\section{Classical lukewarm black holes}

A very convenient representation of the line element describing
the lukewarm black holes in
the Einstein-Maxwell theory with the (positive) cosmological
constant is that in terms of the horizons. Denoting the location 
of the event horizon by $a$ and the location of the cosmological 
horizon by $b$ the line element may be written in the form
\begin{equation}
ds^{2} = -f_{0} (r) dt^{2} + \frac{1}{f_{0}(r)} dr^{2} + r^{2} 
\left(d\theta^{2} + \sin^{2}\theta d\phi^{2} \right),
\label{spec1}
\end{equation}
where
\begin{equation}
f_{0}(r) = \left(1 - \frac{ab}{(a+b)r} \right)^{2} -\frac{r^{2}}{(a+b)^{2}}.
\end{equation}
Such a configuration is allowed provided
\begin{equation}
Q^{2} = \left( \frac{ab}{a+b} \right)^{2} \hspace{0.5cm}{\rm and} \hspace{0.5cm} 
\Lambda = \frac{3}{(a+b)^{2}}.
                    \label{luke}
\end{equation}
It means that the electric charge $Q$ (if there is a magnetic charge, $P,$
$Q^{2}$ must be substituted by $Z^{2} = Q^{2}+P^{2}$) and $\Lambda$
completely determines the lukewarm solution.

The simplest way to demonstrate that the line element of the lukewarm 
Reissner-Nordstr\"om-de Sitter black hole can be expressed solely 
in terms of $a$ and $b$ is to solve  the Einstein-Maxwell equations 
with the cosmological term for a general static and spherically-symmetric 
metric
\begin{equation}
  ds^2 = - A (r) dt^2 + B (r) dr^2 + 
r^2 \left(d\theta^{2} + \sin^{2}\theta d\phi^{2} \right)
\label{familiar}
\end{equation}
with the boundary conditions $B^{- 1} (a) = 0$ and 
$A(b) B (b) = 1.$
Solving the field equations and making use of the boundary conditions yields
\begin{equation}
A(r) =   \frac{1}{B (r)} = 1 - \frac{a}{r} - \frac{Q^2}{r a} + \frac{\Lambda
  a^3}{3 r} + \frac{Q^2}{r^2} - \frac{\Lambda r^2}{3},
\label{aabb}
\end{equation}
which, with the substitution
\begin{equation}
  M_H = \frac{a}{2} - \frac{Q^2}{2 a} + \frac{\Lambda a^2}{6}
\end{equation}
leads to the line element (\ref{familiar}) written in a more familiar form
\begin{equation}
  A (r) = B^{- 1} (r) = 1 - \frac{2 M_H}{r} + \frac{Q^2}{r^2} - \frac{\Lambda
  r^2}{3}.
\end{equation}
We shall refer to $M_H$ as the horizon-defined mass. 
Simple analysis
indicate that equation $A (r) = 0$ can have, depending on the values of the
parameters, three, two or one distinct positive solutions. The above
configurations can, therefore, have three distinct horizons located at zeros
of $A (r),$ a degenerate and a nondegenerate horizon, and, finally, one triply
degenerate horizon. Let us denote remaining solutions of the equation $A (r) = 0$ by
$r_{--}$ and $r_-$ $(r_{--} < r_{-} \leq a \leq b).$
Solving of the system of equations
\begin{equation}
   A (b) = 0 \hspace{0.5cm}{\rm and} \hspace{0.5cm}
  \kappa (a) + \kappa (b) = 0,
\end{equation}
where
\begin{equation}
  \kappa (r_i) = \frac{1}{2} \left(- g_{tt} g_{rr} \right)^{-1 / 2} \frac{d}{dr}
  g_{tt}
\end{equation}
with respect to $\Lambda$ and $Q^{2}$ one obtains (\ref{luke}). Finally, substituting the
thus obtained $\Lambda$ and $Q^{2}$ into Eq.~\ref{aabb} gives (\ref{spec1}).

\section{Semiclassical lukewarm black holes}
\subsection{General equations in $(a,b)$ parametrization }
Now, let us consider the semiclassical Einstein field  equations
\begin{equation}
R_{i}^{j} - \frac{1}{2} R \delta_{i}^{j} + \Lambda \delta_{i}^{j}  =
8\pi \left( T_{i}^{(em)j} + T_{i}^{(q)j} \right),
\label{semiEinstein}
\end{equation}
where $T_{i}^{(em)j}$ is the stress-energy tensor of the electromagnetic
field and $T_{i}^{(q)j}$ is the renormalized stress-energy tensor of
some quantized test field or fields, evaluated in suitable state. 
Optimally, the renormalized stress-energy tensor of the quantized fields
should functionally depend on the metric tensor, or, at least, on the wide class
of geometries. Unfortunately, because of the mathematical complexity 
of the problem one is forced to refer either to the analytic approximations 
or numerics or both. Here we shall use the renormalized stress-energy tensor
of the massive scalar field with an arbitrary curvature coupling in a large 
mass limit. Such a tensor can be calculated from the effective 
action, $W_{R},$ constructed within the framework of the Schwinger-DeWitt method~\cite{Bryce1}. 
Although the results appear to be state-independent the formulae used in this paper
have been constructed with the assumption that the Green functions are defined
in the spaces with the positive-definite metric.
Of course, in this approach one ignores creation of real particles but the analyses 
that have been carried out so far suggest that for sufficiently massive fields
it provides a reasonably accurate approximation~\cite{PaulA}. 

The approximate renormalized one-loop effective action of the quantized massive 
fields in the large mass limit is given by a well-known formula
\begin{equation}
W_{R}\,=\,\frac{1}{32\pi^{2}}\sum_{n=3}^{\infty}
\frac{(n-3)!}{(m^{2})^{n-2}}\int d^{4}x \sqrt{g} [a_{n}],
                           \label{Weff}
\end{equation}
where each $[a_{n}]$ has dimensionality of $[length]^{-2 n}$ and is constructed
from  the Riemann tensor, its covariant derivatives up to $2 n-2$ order and appropriate 
contractions. For the technical details of this approach the reader is referred, for
example, to Refs.~\cite{Barvinsky:1985an,FZ3} and the references cited therein.
Inspection of Eq.~(\ref{Weff}) shows that the lowest term of the approximate $W_{R}$ 
is to be constructed from the (integrated) coincidence limit of the fourth Hadamard-
Minakshisundaram-DeWitt-Seely coefficient, $[a_{3}],$ whereas the next to leading term 
should be constructed form $[a_{4}].$ Here we will confine ourselves to the first term
of the expansion~(\ref{Weff}) and briefly discuss some  general features of the 
second-order term at the end of the paper. The approximate renormalized stress-energy
tensor can be constructed from $W_{R}$ using the standard relation 
\begin{equation}
T^{ab} = \frac{2}{g^{1/2}}\frac{\delta}{\delta g_{ab}} W_{R}.
    \label{def_set}
\end{equation}

It should be noted that even if the renormalized stress-energy tensor
is known the resulting semiclassical field equations are far too complicated
to be solved exactly. Since the quantum effect are expected to be small, 
it is reasonable to assume that the quantum-corrected lukewarm
black hole is described by a set of parameters that are close to their classical 
counterparts. Introducing slightly distorted metric potentials
\begin{equation}
A(r) = f_{0}(r) + \alpha(r)
\label{alfa}
\end{equation}
and
\begin{equation}
1/B(r) = f_{0}(r) + \beta(r),
\label{bett}
\end{equation}
where $\alpha(r)$ and $\beta(r)$ are small corrections to the main
approximation, expanding and retaining only the linear terms in the resulting 
differential equations one obtains  
\begin{equation}
\frac{1}{r} \frac{d\beta(r)}{dr} + \frac{\beta(r)}{r^{2}} -\frac{1}{12\pi m^{2}}t_{t}^{t} = 0
\end{equation}
and
\begin{equation}
\frac{1}{r} \frac{d \alpha(r)}{dr} -2 p(r) \alpha(r) +q(r) \beta(r) -\frac{1}{12\pi m^{2}}t_{r}^{r} =0,
\end{equation}
where
\begin{equation}
p(r) ={\frac {{a}^{2}br-{a}^{2}{b}^{2}+a{b}^{2}r-{r}^{4}}{{r}^{2}
 \left( b-r \right)  \left( a-r \right)  \left( ab-ar-br-{r}^{2}
 \right) }}
\end{equation}
and
\begin{equation}
q(r) = {\frac {{a}^{2}{b}^{2}-{r}^{2}{a}^{2}-2\,{r}^{2}ab+3\,{r}^{4}-{r}^{2}
{b}^{2}}{{r}^{2} \left( b-r \right)  \left( r-a \right)  \left( ab-ar-
br-{r}^{2} \right) }}.
\end{equation}
The general solution to the system can be written
\begin{equation}
\beta(r) = \frac{1}{12\pi m^{2} r}\int t_{t}^{t}(r) r^{2} dr + C_{1}
\label{bbb}
\end{equation}
\begin{equation}
\alpha(r) =-\frac{P(r)}{6\pi m^{2}} \int \frac{1}{P(r)}\left( 12 q(r) \beta(r) \pi m^{2}
- t_{r}^{r}(r) \right)dr +P(r) C_{2}, 
\label{aaa}
\end{equation}
where
\begin{equation}
P(r) = \exp\left(\int p(r) dr \right).
\end{equation}
and, consequently,  the construction of the general solution reduces to
two  quadratures. Here $t_{a}^{b} = 96 \pi^{2} m^{2} T_{a}^{(q)b}.$
In deriving the above (formal) solution we have ignored 
subtleties associated with the stress-energy tensor itself, i.e., it is
tacitly assumed that it is regular on the horizons. Fortunately, it turns out
that the stress-energy tensor considered in this paper is free of 
such complications. The integration constants $C_{1}$ and $C_{2}$
should be determined from physically motivated boundary conditions. 

\subsection{The approximate stress-energy tensor}
\label{sec:sss}
To solve the semiclassical Einstein field equations in a self-consistent
way the renormalized stress-energy tensor describing the quantized
source term is required.  Such a  tensor for a given field should functionally 
depend on the generic metric tensor. Unfortunately, because of the unavoidable
complexities of the problem, its exact form is unknown. Provided we are 
interested in the analytical calculations all we can do is to look for some 
reasonable  approximation~\cite{FZ3,AHS95,Matyjasek:1999an,kocio1,lemosT,kocio:2009,MatryZw}.  

The first-order approximation to the renormalized effective action of the quantized 
massive scalar  field with arbitrary coupling to the curvature  $\xi$ satisfying 
the covariant Klein-Gordon equation
\begin{equation}
     \left( \Box \,-\,\xi R\,-\,m^{2}\right) \phi \,=\,0,  
                                  \label{wave}
\end{equation}
can be constructed from the coincidence limit of the coefficient  
\begin{equation}
[a_{3}] = a_{3}^{(0)} + \xi a_{3}^{(1)} + \xi^{2} a_{3}^{(2)} + \xi^3 a_{3}^{(3)},
                                           \label{a3a}
\end{equation}
where
\begin{eqnarray}
a_{3}^{(0)} &=& \frac{11}{1680}  R^3+
\frac{17}{5040}  R_{;a}^{\pp{;\pp{a}}}  R_{\pp{;\pp{a}}}^{;a}-
\frac{1}{2520}  R_{ab;c}^{\pp{a}\pp{b}\pp{;\pp{c}}}  R_{\pp{a}\pp{b}\pp{;\pp{c}}}^{ab;c}-
\frac{1}{1260}  R_{ab;c}^{\pp{a}\pp{b}\pp{;\pp{c}}}  R_{\pp{a}\pp{c}\pp{;\pp{b}}}^{ac;b}\nonumber \\ 
 &+
&\frac{1}{560}  R_{abcd;e}^{\pp{a}\pp{b}\pp{c}\pp{d}\pp{;\pp{e}}}  
R_{\pp{a}\pp{b}\pp{c}\pp{d}\pp{;\pp{e}}}^{abcd;e}+
\frac{1}{180}  R  R_{;a\pp{a}}^{\pp{;\pp{a}}a}+
\frac{1}{280}  R_{;a\pp{a}b\pp{b}}^{\pp{;\pp{a}}a\pp{b}b}+
\frac{1}{420}  R_{;ab}^{\pp{;\pp{a}}\pp{b}}  R_{\pp{a}\pp{b}}^{ab}\nonumber \\ 
 &-
&\frac{1}{630}  R_{ab;c\pp{c}}^{\pp{a}\pp{b}\pp{;\pp{c}}c}  R_{\pp{a}\pp{b}}^{ab}-
\frac{109}{2520}  R  R_{ab}^{\pp{a}\pp{b}}  R_{\pp{a}\pp{b}}^{ab}+
\frac{73}{1890}  R_{ab}^{\pp{a}\pp{b}}  R_{c\pp{a}}^{\pp{c}a}  R_{\pp{b}\pp{c}}^{bc}+
\frac{1}{210}  R  R_{abcd}^{\pp{a}\pp{b}\pp{c}\pp{d}}  R_{\pp{a}\pp{b}\pp{c}\pp{d}}^{abcd}\nonumber \\ 
 &+
&\frac{1}{105}  R_{ab;cd}^{\pp{a}\pp{b}\pp{;\pp{c}}\pp{d}}  R_{\pp{a}\pp{c}\pp{b}\pp{d}}^{acbd}+
\frac{19}{630}  R_{ab}^{\pp{a}\pp{b}}  R_{cd}^{\pp{c}\pp{d}}  R_{\pp{a}\pp{c}\pp{b}\pp{d}}^{acbd}-
\frac{1}{189}  R_{abcd}^{\pp{a}\pp{b}\pp{c}\pp{d}}  R_{ef\pp{a}\pp{b}}^{\pp{e}\pp{f}ab}  
R_{\pp{c}\pp{d}\pp{e}\pp{f}}^{cdef},
                                                          \label{a3b}
\end{eqnarray}

\begin{eqnarray}
a_{3}^{(1)} &=& -
 \frac{1}{72} R^3-
\frac{1}{30}  R_{;a}^{\pp{;\pp{a}}}  R_{\pp{;\pp{a}}}^{;a}-
\frac{11}{180}  R  R_{;a\pp{a}}^{\pp{;\pp{a}}a} -
\frac{1}{180}  R  R_{abcd}^{\pp{a}\pp{b}\pp{c}\pp{d}}  R_{\pp{a}\pp{b}\pp{c}\pp{d}}^{abcd}
  \nonumber \\ 
 &-
&\frac{1}{60}  R_{;a\pp{a}b\pp{b}}^{\pp{;\pp{a}}a\pp{b}b}-
\frac{1}{90}  R_{;ab}^{\pp{;\pp{a}}\pp{b}}  R_{\pp{a}\pp{b}}^{ab}+
\frac{1}{180}  R  R_{ab}^{\pp{a}\pp{b}}  R_{\pp{a}\pp{b}}^{ab},
                                                        \label{a3c}
\end{eqnarray}

\begin{equation}
a_{3}^{(2)} = \frac{1}{12}R^3 +
\frac{1}{12}  R_{;a}^{\pp{;\pp{a}}}  R_{\pp{;\pp{a}}}^{;a}+
\frac{1}{6}  R  R_{;a\pp{a}}^{\pp{;\pp{a}}a},
                                                         \label{a3d}
\end{equation}
and
\begin{equation}
a_{3}^{(3)} = -
 \frac{1}{6} R^3
                                                        \label{a3e}
\end{equation}
It is evident that the approximate stress-energy tensor of the quantized 
massive field constructed from the effective action $W_{R}$  depends 
functionally on the  metric as it is constructed solely from the Riemann tensor,
its contractions and covariant derivatives to some required order. The type of 
the field enters the general formulas by the spin-dependent numeric 
coefficients. There is no need, when computing the stress-energy tensor, to 
retain all terms in $[a_{3}].$ Indeed, the total divergences can be discarded and 
further simplifications of the effective action are possible. Here we display 
the coefficient $[a_{3}(x,x')]$  in its full form simply because it will be 
of use in the calculations of the field fluctuation. 

Now, repeating the steps 
of Refs~\cite{Matyjasek:1999an,kocio1} (to which the reader is referred for 
additional informations), after some algebra, one obtains the covariantly 
conserved tensor (\ref{def_set}), with the tensor $t_{a}^{b}$  is given by
\begin{eqnarray}
t_{t}^{t}& =& -\frac{2327 a^6 b^6}{105 r^{12} (a+b)^6}+\frac{2452 a^5 b^5}{35
   r^{11} (a+b)^5}-\frac{8611 a^4 b^4}{105 r^{10}
   (a+b)^4}+\frac{643 a^4 b^4}{35 r^8 (a+b)^6}+\frac{4442 a^3
   b^3}{105 r^9 (a+b)^3}\nonumber \\
&&-\frac{2468 a^3 b^3}{105 r^7
   (a+b)^5}-\frac{57 a^2 b^2}{7 r^8 (a+b)^2}+\frac{289 a^2
   b^2}{35 r^6 (a+b)^4}
+\frac{a^2 b^2}{15 r^4 (a+b)^6}
+\xi^3 \left(\frac{432}{(a+b)^6}-\frac{432
   a^2 b^2}{r^4 (a+b)^6}\right) \nonumber \\
&&-\frac{37}{21 (a+b)^6} +\xi^2 \left(\frac{216 a^2 b^2}{r^4
   (a+b)^6}-\frac{216}{(a+b)^6}\right)
+\xi \left(\frac{546 a^6 b^6}{5 r^{12}
   (a+b)^6}-\frac{1808 a^5 b^5}{5 r^{11} (a+b)^5}+\frac{6664 a^4
   b^4}{15 r^{10} (a+b)^4}\right.\nonumber \\
&&\left.
-\frac{462 a^4 b^4}{5 r^8
   (a+b)^6}-\frac{240 a^3 b^3}{r^9 (a+b)^3}+\frac{656 a^3 b^3}{5
   r^7 (a+b)^5}+\frac{48 a^2 b^2}{r^8 (a+b)^2}-\frac{48 a^2
   b^2}{r^6 (a+b)^4}-\frac{26 a^2 b^2}{r^4 (a+b)^6}+\frac{174}{5
   (a+b)^6}\right),
                                         \label{czasowa}
\end{eqnarray}
\begin{eqnarray}
t_{r}^{r} &=&\frac{421 a^6 b^6}{105 r^{12} (a+b)^6}-\frac{52 a^5 b^5}{3 r^{11} (a+b)^5}
+\frac{949 a^4 b^4}{35 r^{10} (a+b)^4}-\frac{29 a^4 b^4}{3 r^8
   (a+b)^6}-\frac{646 a^3 b^3}{35 r^9 (a+b)^3}
-\frac{37}{21 (a+b)^6}
 \nonumber \\
&&+\frac{1604 a^3 b^3}{105 r^7 (a+b)^5}
+\frac{33 a^2 b^2}{7 r^8 (a+b)^2}-\frac{97 a^2 b^2}{15 r^6
   (a+b)^4}+\xi^3 \left(\frac{432}{(a+b)^6}
-\frac{432 a^2 b^2}{r^4 (a+b)^6}\right)
+\frac{29 a^2 b^2}{15 r^4 (a+b)^6}
\nonumber \\
&&+\xi^2 \left(\frac{216 a^2 b^2}{r^4
   (a+b)^6}-\frac{216}{(a+b)^6}\right)
+\xi \left(-\frac{78 a^6 b^6}{5 r^{12} (a+b)^6}+\frac{336 a^5 b^5}{5
   r^{11} (a+b)^5} 
 -\frac{1592 a^4 b^4}{15 r^{10} (a+b)^4}
+\frac{42 a^4 b^4}{r^8 (a+b)^6}
\right. \nonumber \\
&&\left.
+\frac{368 a^3 b^3}{5 r^9 (a+b)^3}
-\frac{336 a^3 b^3}{5
   r^7 (a+b)^5}-\frac{96 a^2 b^2}{5 r^8 (a+b)^2}
+\frac{144 a^2 b^2}{5 r^6 (a+b)^4}-\frac{178 a^2 b^2}{5 r^4 (a+b)^6}
+\frac{174}{5
   (a+b)^6}\right)
                                         \label{radialna}
\end{eqnarray}
and
\begin{eqnarray}
t_{\theta}^{\theta} &=& -\frac{497 a^6 b^6}{15 r^{12} (a+b)^6}
+\frac{11404 a^5 b^5}{105 r^{11} (a+b)^5}
-\frac{13903 a^4 b^4}{105 r^{10} (a+b)^4}+\frac{1769 a^4 b^4}{105
   r^8 (a+b)^6}+\frac{2486 a^3 b^3}{35 r^9 (a+b)^3}
-\frac{a^2 b^2}{r^4 (a+b)^6}
\nonumber \\
&&
-\frac{108 a^3 b^3}{5 r^7 (a+b)^5}-\frac{99 a^2 b^2}{7 r^8 (a+b)^2}+\frac{683 a^2 b^2}{105
   r^6 (a+b)^4}   -\frac{37}{21 (a+b)^6} +  \xi^3 \left(\frac{432 a^2 b^2}{r^4 (a+b)^6}
+\frac{432}{(a+b)^6}\right)
\nonumber \\
&&
+\xi^2 \left(-\frac{216 a^2 b^2}{r^4
   (a+b)^6}-\frac{216}{(a+b)^6}\right)
+\xi \left(\frac{702 a^6 b^6}{5 r^{12} (a+b)^6}-\frac{2272 a^5 b^5}{5 r^{11}
   (a+b)^5}+\frac{8216 a^4 b^4}{15 r^{10} (a+b)^4}
\right. \nonumber \\
&& \left.
-\frac{342 a^4 b^4}{5 r^8 (a+b)^6}
-\frac{1456 a^3 b^3}{5 r^9 (a+b)^3}+\frac{416 a^3 b^3}{5 r^7
   (a+b)^5}+\frac{288 a^2 b^2}{5 r^8 (a+b)^2}-\frac{24 a^2 b^2}{r^6 (a+b)^4}
+\frac{154 a^2 b^2}{5 r^4 (a+b)^6}+\frac{174}{5
   (a+b)^6}\right).
                                   \label{katowa}
\end{eqnarray}
These components can also be constructed from the Euler-Lagrange 
equations~\cite{MatryZw} with the Lagrangian depending on the time and radial
components of the metric tensor, their derivatives and coordinate $r.$  
Because of the complexity of the calculations  of the stress-energy tensor this 
alternative approach  may serve as a useful check.

In order to establish the regularity of the stress-energy tensor at the
horizon it is necessary to transform it into the coordinate system that is
regular there. Alternatively, one can use in this regard a freely falling
frame. For radial motion, the orthogonal vectors of the frame are the unit
tangent to the geodesic $e_{(0)}^i$ and the three spacelike mutually
perpendicular unit vectors $e_{(j)}^i .$ Integrating the geodesic equation one
obtains
\begin{equation}
  u^a = \left[ \frac{C}{A}, - \sqrt{\frac{1}{B} \left( \frac{C^2}{A} - 1
  \right)}, 0, 0 \right]
\end{equation}
and
\begin{equation}
  n^a = \left[ \pm \frac{\sqrt{C^2 - A}}{A}, \mp \frac{C}{\sqrt{AB}}, 0, 0
  \right],
\end{equation}
where $C$ is the energy per unit mass along the geodesic. Elementary manipulations
shows that the components $\tT_{(0) (0)},$ $\tT_{(1) (1)}$ and $\tT_{(0) (1)}$ 
of the stress-energy tensor in a freely falling frame are given by
\begin{equation}
  \tT_{(0) (0)} = \frac{C^2}{A} \left( T_r^r - T_t^t \right) - T_r^r,
\end{equation}
\begin{equation}
  \tT_{(1) (1)} = \frac{C^2}{A} \left( T_r^r - T_t^t \right) + T_r^r
\end{equation}
and
\begin{equation}
  \tT_{(0) (1)} = \tT_{(1) (0)} = \frac{C \sqrt{C^2 - A}}{A} \left( T_r^r -
  T_t^t \right).
\end{equation}
Since the difference between the radial and time components of the
stress-energy tensor (\ref{czasowa}-\ref{katowa}) factors as
\begin{equation}
  T_r^{(q)r} - T_t^{(q)t} = A (r) F (r),
\end{equation}
where $F (r)$ is a regular function, one concludes that the stress-energy
tensor in a  frame freely falling from the cosmological horizon or falling 
on the event horizon is regular in a physical sense.

\subsection{Semiclassical lukewarm black holes in $(a,b)$ parametrization}
As is seen from Eqs.~(\ref{bbb}) and (\ref{aaa}) the  general solution of the 
linearized system (\ref{semiEinstein}) requires two simple quadratures,  which, 
after some calculations, yield rather long expressions. The corrections to the
``metric potentials''  $1/B(r)$ and $A(r),$ i.e., the functions  
$\beta(r)$ and $\alpha(r)$ have the structure 
\begin{equation}
\beta(r) =\frac{1}{\pi m^{2} (a+b)^{6}} \sum_{i=0}^{3} \beta_{i}(r) \xi^{i} 
+ \frac{C_{1}}{r}
\label{bet}
\end{equation}
and 
\begin{equation}
\alpha(r) = \beta(r) + \frac{f_{0}(r)}{\pi m^{2}(a+b)^{4}} \left[ \sum_{i=0}^{1} \alpha_{i}(r) \xi^{i}
-6 \xi^{2} + 12 \xi^{3} \right] - (a+b)^{2} C_{2},
\label{al}
\end{equation}
where $f_{0}(r)$ is given by (\ref{spec1}) and the exact form of the functions 
$\alpha_{i}(r)$ and $\beta_{i}(r)$  are relegated to the appendix.
Of the two integration constants, only $C_{1}$ affects location
of the horizons; the second integration constant is left free throughout  
the calculation. If it exists, the quantum-corrected lukewarm black hole 
must satisfy the same requirements as its classical counterpart, and, consequently,
in order to determine the line element describing such a configuration,
one has to solve the system of algebraic equations:
\begin{equation}
A(r_{+}) = A(r_{c}) =0
\end{equation}
and
\begin{equation}
\kappa(r_{+}) + \kappa(r_{c}) =0,
\end{equation}
where $r_{+} = a + r^{(1)}_{+} $ and $r_{c} = b + r^{(1)}_{c},$
with respect to $C_{1},$ $r^{(1)}_{+} $ and $r^{(1)}_{c}.$  Here $\rp$ and $r_{c}$
correspond respectively to the event horizon and the cosmological horizon,
whereas $r^{(1)}_{+}$ and $ r^{(1)}_{c}$ are small corrections.
Now, simple manipulations give
\begin{equation}
r_{+}^{(1)} = \frac{1}{\pi m^{2} (a-b)(a+b)^{5}}\left(W_{+}^{(0)} + W_{+}^{(1)} \xi\right) 
+ \frac{1}{\pi m^{2} (a-b)(a+b)^{4}}\left(W_{+}^{(2)} \xi^{2} +W_{+}^{(3)}\xi^{3}\right) 
\label{event}
\end{equation}
and
\begin{equation}
C_{1} =  \frac{1}{\pi m^{2} (a+b)^{7}}\left(V^{(0)} + V^{(1)} \xi\right) 
+ \frac{1}{\pi m^{2} (a+b)^{5}}\left(V^{(2)} \xi^{2} +V^{(3)}\xi^{3}\right) 
\label{ce1}
\end{equation}
where functions $W_{+}^{(k)}$ and $V^{(k)}$ are listed in Appendix.
It should be noted that 
\begin{equation}
r_{c}^{(1)}(a,b) = -r_{+}^{(1)}(a\to -b, b\to -a).
\label{cosmoH}
\end{equation}
i.e., the correction to the cosmological horizon equals
minus the correction to the event horizon with simultaneous 
substitution $a \to - b$ and $b \to -a.$ 
The integration constant $C_{2}$ can be determined using, for example,
the condition $A(r_{c}) B(r_{c}) = 1.$

The equations (\ref{bet}) with (\ref{ce1}) and (\ref{al}) solve the problem 
completely: the semiclassical lukewarm configuration is characterized 
by the cosmological constant and (total) charge as given by (\ref{luke}), 
whereas the quantum corrections to the event and cosmological horizon are 
given by (\ref{event}) and (\ref{cosmoH}), respectively.

\subsection{Semiclassical black holes in $(\rp,\rc)$ parametrization}

Now we shall show that the above procedure is equivalent 
to a more familiar approach, in which one looks for a lukewarm 
solution parametrized by the exact location of the event horizon and
the zeroth-order approximation to the cosmological horizon.
First, let us assume that the cosmological constant is a parameter in a space 
of theories rather than the space of solutions. Now, one can solve the 
semiclassical Einstein field equations for a line element of the form 
(\ref{familiar}) with
\begin{equation}
A(r) = \left(1-\frac{2 M(r)}{r} \right) e^{2\psi(r)}
\end{equation}
and
\begin{equation}
B(r) = \left(1-\frac{2 M(r)}{r} \right)^{-1},
\end{equation}
where  $M(r) = M_{0}(r) + M_{1}(r)$ and $\psi(r) = \psi_{0}(r) + \psi_{1}(r)$
and the functions $M_{1}(r)$ and $\psi_{1}(r)$ are small corrections to the
main approximation.
In constructing the linearized solution we adopt the natural conditions
$M_{0}(r_{+}) = r_{+}/2,$ $M_{1}(r_{+}) =0$ and $\psi_{0}(r) =0,$ leaving 
unspecified the integration constant, say $\tilde{C}_{2},$  which appears 
as a result of integration 
of the differential equation for $\psi_{1}(r).$  The zeroth-order solution 
is therefore parametrized by the exact location of the event horizon, $r_{+},$ 
and the  charge $Q^{2}.$  The  lukewarm configuration can be expressed 
in terms of $\rp$ and the zeroth-order approximation to the cosmological 
horizon $\rc$
\begin{equation}
  A (r) = B^{- 1} (r) = \left( 1 - \frac{\rp \rc}{\left( \rp
  + \rc \right) r} \right)^2 - \frac{r^2}{\left( \rp + \rc
  \right)^2}.
\end{equation} 
In this approach we do not attribute any physical significance to the
zeroth-order solution. Once again, the first-order corrections can 
easily be constructed by the two simple quadratures and the quantum-corrected
lukewarm black hole is characterized by the exact location of the event 
horizon, $\rp,$ the location of the cosmological horizon
\begin{equation}
r_{c} = \rc + \delta(\rp,\rc)
\end{equation}
and the relation between the (total) charge and $\rp$ and $\rc:$
\begin{equation}
Q^{2} = \left(\frac{\rp\rc}{\rp+\rc} \right)^{2} + \Delta(\rp,\rc)
\end{equation}
By the assumption we made about the cosmological constant it is still given by
\begin{equation}
\Lambda = \frac{3}{(\rp+\rc)^2}.
\end{equation} 
Although we have calculated both $\delta(\rp,\rc)$ and $\Delta(\rp,\rc)$
we shall not display them here, simply because they are rather lengthy. Moreover,
they can easily be constructed form the formulas listed in the appendix by switching
from $(a,b)$ representation to $(\rp,\rc),$ (i. e., by the reverse procedure 
to that discussed below).

For the same black hole configuration one can switch from $(\rp,\rc)$ 
representation to $(a,b)$ making use
of the equations
\begin{equation}
\frac{3}{(\rp+\rc)^2} = \frac{3}{(a+b)^{2}}
\label{repr1}
\end{equation}
and
\begin{equation}
 \left(\frac{\rp\rc}{\rp+\rc} \right)^{2} + \Delta(\rp,\rc)
= \left( \frac{ab}{a+b} \right)^{2}
\label{repr2}
\end{equation}
Indeed, substituting
\begin{equation}
\rp = a + a_{1}
\hspace{0.5cm}{\rm and} \hspace{0.5cm}
\rc = b+ b_{1},
\end{equation}
where $a_{1}$ and $b_{1}$ are small corrections, into the 
system (\ref{repr1}) and (\ref{repr2}) one obtains 
\begin{equation}
a_{1} = - b_{1} = \frac{(a+b)^{2} }{2 a b (a-b)} \Delta(a,b).
\end{equation}
Since we are interested in the first-order calculations, we can safely change
the arguments of the function $ \Delta(\rp,\rc).$ 
It can be demonstrated that $r_{+}$ 
and $r_{c} =b + b_{1} + \delta(a,b)$ are given respectively by
by (\ref{event}) and (\ref{cosmoH}), as expected. 
Moreover,  both approaches yield identical results for the metric
tensor, provided the integration constant $\tilde{C}_{2}$ is 
related to the constant $C_{2}$ by
\begin{equation}
\tilde{C}_{2} =  - (a+b)^2 C_{2} -\frac{1}{\pi m^{2} (a+b)^{4}} \left( 
\frac{37}{756} -\frac{29}{30} \xi +6 \xi^{2} -12 \xi^{3} 
\right).
\end{equation}
As before, to determine $C_{2}$ (and hence $\tilde{C}_{2}$) additional 
informations are required. 

\section{Final remarks}
In this paper we have constructed perturbative solutions
to the semiclassical Einstein field equations describing 
spherically-symmetric and static lukewarm black 
hole. The total source term is composed of two parts:
the (classical) stress-energy tensor of the electromagnetic field and 
the renormalized stress-energy tensor of the quantized massive 
scalar field in a large mass limit. 
In the course of our calculations we used two different parametrizations, 
and, assuming that the first-order results describe the same black hole 
configuration we constructed the transformation rules from the one parametrization 
to the other.  
In the parametrization $(a,b)$ we first calculated the zeroth-order solution.
Subsequently, making use of the quantum part of the total stress-energy tensor
constructed in the classical background we calculated the corrections to the
metric potentials and the corrections to the horizons. This procedure 
can be thought of as switching the quantized field on and analyzing its
influence on the classical background via the back-reaction.
On the other hand, in the parametrization $(\rp,\rc),$ we are looking for
a self-consistent solution from the very beginning. This requires a generic
tensor which depend functionally on the metric tensor. Since the calculations
in that case are rather involved and produce complex results, we discussed them 
only briefly. 

We conclude this paper with a number of comments:
\begin{enumerate}
\item Here we have considered only the quantized massive 
scalar fields with $\xi R \phi$ coupling in a large mass
limit. 
Since the approximate stress energy tensors of the massive spinor
and the massive vector fields in for a generic metric are known, 
the results presented here can easily be extended to these cases.

\item Once the renormalized stress-energy tensor is known
a similar analysis can be carried out for the quantized massless fields.  
Unfortunately, although the renormalized stress-energy tensor
of various fields are well documented in the Schwarzschild 
spacetime (see for example Refs.~\cite{Page:1982fm,Brown:1985ri,Brown:1986jy,Frolov:1987gw,
Jirinek96prd,Jirinek97cqg,Jirinek97prd,Jirinek98prd,Groves:2002mh,Carlson:2003ub} 
and the references cited therein)
less is known about $T_{a}^{b}$ in  more complicated geometries.
The remarks made in subsection~\ref{sec:sss} remain intact and it
is crucial to check the regularity of the stress-energy tensor 
on the horizons.

\item Since the stress-energy tensor is constructed solely
form the Riemann tensor, its derivatives and the metric 
one expects that similar calculations can be performed
in other theories in which the higher curvature terms
appear. For example, although we have considered only the main approximation
to the stress-energy tensor constructed from the integrated coincidence
limit of the Hadamard-DeWitt coefficient  $a_{3}(x,x')$
the calculations can be extended, at the price of 
the technical complications, to the next-to-leading term
involving functional derivatives of the coincidence limit of the 
coefficient $a_{4}(x,x').$ Preliminary calculations indicate 
that it is possible to construct the lukewarm black hole 
in such a case.

\item It could be shown that the approximation to the mean value of the field 
fluctuation in a large mass limit is given by~\cite{Frolov_hab}
\begin{equation}
\langle \phi^{2}\rangle = \frac{1}{16\pi^{2}}\sum_{n=2}^{N}\frac{(n-2)!}{m^{2(n-1)}} [a_{n}],
\end{equation}
where $N-1$ is the number of terms retained in the expansion.
Taking 
\begin{equation}
[a_{2}] = \frac{1}{6}\left(\frac{1}{5} -\xi \right) \Box R +
 \frac{1}{2} \left(\frac{1}{6} -\xi \right)^{2}R^{2} 
-\frac{1}{180} R_{ab}R^{ab}+
\frac{1}{180} R_{abcd} R^{abcd}
\end{equation} and $[a_{3}]$ as given by (\ref{a3a}-\ref{a3e}) one can 
calculate the first two terms of the approximation. For example,  routine
calculations carried out in the spacetime of the 
Einstein-Maxwell lukewarm black hole (\ref{spec1}) give the for the 
main approximation
\begin{equation}
16 \pi^{2} m^{2} \langle \phi^{2}\rangle  = \frac{4}{15}\frac{a^{2} b^{2}}{r^{6} w^{2}} 
-\frac{8}{15} \frac{a^{3}b^{3}}{r^{7} w^{3}}
+ \frac{13}{45} \frac{a^{4} b^{4}}{r^{8} w^{4}}
+\frac{29}{15 w^{4}} -\frac{24}{w^{4}}\xi + \frac{72}{w^{4}}\xi^{2},
\end{equation}
where $w = a + b.$ 
\item Since the cosmological ``constant'' is expected to vary in time (see for 
example~\cite{Padmana,Nov,Mat} and the references cited therein)
it would be interesting to analyze the charged black holes in such a dynamic 
environment. This, however, would require a  deeper understanding of quantum phenomena
taking place in the spacetime of nonstatic black holes.    
\item Extension of the results presented in this paper to the black hole spacetimes
of $d-$dimensions requires detailed knowledge of the higher HMDS coefficients.
\end{enumerate} 
Some of the listed problems are under active investigations and the results will be 
published elsewhere.
\appendix*
\section{}
The geometry of the semiclassical lukewarm black hole in the $(a,b)$ parametrization
is described by the line element (\ref{familiar}) with (\ref{alfa}) and (\ref{bett}).
 The $\beta(r)$ is given by 
\begin{equation}
\beta(r) =\frac{1}{\pi m^{2} (a+b)^{6}} \sum_{i=0}^{3} \beta_{i}(r) \xi^{i} + \frac{C_{1}}{r}
\end{equation}
where
\begin{eqnarray}
\beta_{0}(r) &=&\frac{2327 a^6 b^6}{11340 r^{10}}-\frac{613 a^6 b^5}{840 r^9}
+\frac{8611 a^6 b^4}{8820 r^8}-\frac{2221 a^6 b^3}{3780 r^7}
+\frac{19 a^6 b^2}{140 r^6}-\frac{613 a^5 b^6}{840 r^9}+\frac{8611 a^5 b^5}{4410 r^8}
-\frac{2221 a^5 b^4}{1260 r^7}+\frac{19 a^5 b^3}{35 r^6}
\nonumber \\
&&+\frac{8611 a^4 b^6}{8820 r^8}
-\frac{2221 a^4 b^5}{1260 r^7}+\frac{1067 a^4 b^4}{2100 r^6}
+\frac{617 a^4 b^3}{1260 r^5}-\frac{289 a^4 b^2}{1260 r^4}
-\frac{2221 a^3 b^6}{3780 r^7}+\frac{19 a^3 b^5}{35 r^6}+\frac{617 a^3 b^4}{1260 r^5}
-\frac{289 a^3 b^3}{630 r^4}
\nonumber \\
&&+\frac{19 a^2 b^6}{140 r^6}-\frac{289 a^2 b^4}{1260 r^4}
-\frac{a^2 b^2}{180 r^2}-\frac{37 r^2}{756},
\end{eqnarray}

\begin{eqnarray}
\beta_{1}(r) &=& -\frac{91 a^6 b^6}{90 r^{10}}+\frac{113 a^6 b^5}{30 r^9}-
\frac{238 a^6 b^4}{45 r^8}+\frac{10 a^6 b^3}{3 r^7}-\frac{4 a^6 b^2}{5 r^6}
+\frac{113 a^5 b^6}{30 r^9}-\frac{476 a^5 b^5}{45 r^8}+\frac{10 a^5 b^4}{r^7}
-\frac{16 a^5 b^3}{5 r^6}
\nonumber \\
&&-\frac{238 a^4 b^6}{45 r^8}+\frac{10 a^4 b^5}{r^7}
-\frac{163 a^4 b^4}{50 r^6}-\frac{41 a^4 b^3}{15 r^5}+\frac{4 a^4 b^2}{3 r^4}
+\frac{10 a^3 b^6}{3 r^7}-\frac{16 a^3 b^5}{5 r^6}-\frac{41 a^3 b^4}{15 r^5}
+\frac{8 a^3 b^3}{3 r^4}
\nonumber \\
&&-\frac{4 a^2 b^6}{5 r^6}+\frac{4 a^2 b^4}{3 r^4}
+\frac{13 a^2 b^2}{6 r^2}+\frac{29 r^2}{30}
\end{eqnarray}
and
\begin{equation}
\beta_{2}(r) = -\frac{\beta_{3}(r)}{2} = -\frac{18 a^2 b^2}{r^2}-6 r^2.
\end{equation}

The function $\alpha(r)$ is given by
\begin{equation}
\alpha(r) = \beta(r) + \frac{f_{0}(r)}{\pi m^{2}(a+b)^{4}} \left[ \sum_{i=0}^{1} \alpha_{i}(r) \xi^{i}
-6 \xi^{2} + 12 \xi^{3} \right] - (a+b)^{2} C_{2},
\end{equation}
where
\begin{equation}
\alpha_{0}(r) =-\frac{229 a^4 b^4}{840 r^8}+\frac{184 a^4 b^3}{441 r^7}
-\frac{5 a^4 b^2}{28 r^6}+\frac{184 a^3 b^4}{441 r^7}-\frac{5 a^3 b^3}{14
   r^6}-\frac{5 a^2 b^4}{28 r^6}+\frac{7 a^2 b^2}{180 r^4}-\frac{37}{756}
\end{equation}
and
\begin{equation}
\alpha_{1}(r) =\frac{13 a^4 b^4}{10 r^8}-\frac{32 a^4 b^3}{15 r^7}
+\frac{14 a^4 b^2}{15 r^6}-\frac{32 a^3 b^4}{15 r^7}+\frac{28 a^3 b^3}{15 r^6}
+\frac{14 a^2 b^4}{15 r^6}-\frac{a^2 b^2}{5 r^4}+\frac{29}{30}.
\end{equation}

The location of the event horizon of the quantum-corrected lukewarm black hole
is given by (\ref{event}) with 
\begin{equation}
W^{(0)}_{+} = \frac{17 a^6}{317520 b^3}+\frac{17 a^5}{31752 b^2}+\frac{41 a^4}{4900 b}
+\frac{b^6}{504 a^3}+\frac{2641 a^3}{396900}+\frac{31 b^5}{35280 a^2}
+\frac{719 a^2 b}{11340}+\frac{173 b^4}{11340 a}+\frac{1208 a b^2}{11025}
+\frac{89 b^3}{2835},
\end{equation}
\begin{equation}
W^{(1)}_{+} = -\frac{a^4}{50 b}-\frac{b^6}{180 a^3}+\frac{109 a^3}{300}-a^2 b
-\frac{11 b^4}{180 a}-\frac{473 a b^2}{300}-\frac{b^3}{10}
\end{equation}
and
\begin{equation}
W_{+}^{2} = - \frac{W_{+}^{(3)}}{2} = -3 a^{2} + 9 a b.
\end{equation}
The integration constant $C_{1}$ is given by (\ref{ce1}) with 
\begin{eqnarray}
V^{(0)} &=& \frac{17 a^7}{158760 b^3}+\frac{17 a^6}{15876 b^2}
+\frac{41 a^5}{2450 b}+\frac{24707 a^4}{396900}+\frac{17 b^7}{158760 a^3}
+\frac{1993 a^3 b}{11340}
\nonumber \\
&&+\frac{17 b^6}{15876 a^2}
+\frac{14039 a^2 b^2}{44100}+\frac{41 b^5}{2450 a}
+\frac{1993 a b^3}{11340}+\frac{24707 b^4}{396900}, 
\end{eqnarray}
\begin{equation}
V^{(1)} = -\frac{a^5}{25 b}-\frac{6 a^4}{25}-\frac{89 a^3 b}{30}
-\frac{439 a^2 b^2}{75}-\frac{b^5}{25 a}-\frac{89 a b^3}{30}
-\frac{6 b^4}{25}
\end{equation}
and
\begin{equation}
V^{(2)} = -\frac{V^{(3)}}{2} = 18 a b.
\end{equation}

\end{document}